\documentclass[12pt]{article}
\usepackage{amssymb}
\usepackage[dvips]{epsfig}
\usepackage{subfigure}

\begin{document}

\author{$^{(a)}$ A. de Souza Dutra\thanks{%
E-mail: dutra@feg.unesp.br} \, and $^{(b)}$ J. A. de Oliveira \\
(a) UNESP Univ Estadual Paulista - Campus de Guaratinguet\'a - DFQ\\
Av. Dr. Ariberto Pereira Cunha, 333
12516-410 Guaratinguet\'{a}, SP, Brazil\\
(b) UNESP Univ Estadual Paulista - IGCE - DEMAC\\
Av. 24A 1515, 13506-700 Rio Claro, SP, Brazil}
\title{{\LARGE Classes of bidimensional isospectral position-dependent mass
systems}} \maketitle

\begin{abstract}
In this work we construct a general class of exactly solvable
non-relativistic bi-dimensional quantum systems with
position-dependent masses (PDM). These systems are isospectral to
a given system with constant mass. The case of a charged particle
with a PDM interacting with an external magnetic field is included
in the present investigation. We apply the approach in order to
construct the SU(2) coherent states in some examples which are
isospectral to the two-dimensional anisotropic harmonic
oscillator, and discuss the impact of the introduction of special
non-homogeneous external magnetic fields.
\end{abstract}

\bigskip

\newpage

\section{Introduction}

The interest in the problem of position-dependent mass (PDM)
systems has been growing in the last years, both from the
non-relativistic and the relativistic point of view
\cite{juliano2}-\cite{epl05}. In fact, there are many applications
to different problems of physics like quantum dots
\cite{dot1,dot2}, compositional graded crystals \cite{crystal},
quantum liquids \cite{liq}, metal clusters \cite{metal}, neutron
stars \cite{stars}, among others. In particular, recently the
Wigner function for some classes of position-dependent
Schroedinger equations were constructed and analyzed in the
one-dimensional case \cite{juliano1}. The approach used in that
work is capable to generate a class of position-dependent mass
systems, which is isospectral to a given exactly solvable
potential with constant mass. In this work, we will extend that
approach to higher dimensions. Moreover, as far as we know, the
majority part of the works dedicated to the research of PDM
systems \cite{juliano2}, deals with one-dimensional systems.
Although, there are physical systems like those where a magnetic
field \cite{pra89} is present, which leads naturally to the need
of a two-dimensional analysis. Finally, to our knowledge, no work
in this subject has discussed the case of PDM in the presence of
magnetic fields. In this work, we intend to partially fill this
gap.

The method will be applied to the case of the anisotropic
two-dimensional harmonic oscillator as well as in the case of
those systems under the influence of some non-homogenous external
magnetic fields. The $SU(2)$ coherent states will be constructed
for a number of systems which we choose to illustrate our results.

In order to get the exact solutions for the two-dimensional
Schroedinger equation we trace two routes. We begin by performing
a general spatial variables change in the case of systems which
are not under the effect of magnetic fields. Then, in the case of
magnetic interaction, we begin by performing a time-dependent
variable transformation followed by the spatial transformation. In
the first case, we consider three examples of real variable
transformations, which are respectively: the so-called polynomial
one, the one using elliptic cylindrical coordinates and, finally,
the bipolar coordinates transformation. Once we have the exact
solutions for the eigenstates of the Schroedinger equation, we
proceed with the construction of the $SU(2)$ coherent states
\cite{juliano2,1Chen2003,3Chen2003,Chen2004}. These last allow us
to acquire a notion of the respective classical behavior of those
position-dependent massive particles, similarly to what happens
with the Wigner functions in one spatial dimension
\cite{juliano1}.

This work is organized as follows: In the section 2 we present the
approach we are going to use. Then, in the section 3 we apply it
to some special mass dependencies, constructing classes of
two-dimensional PDM systems which are isospectral to the
anisotropic bi-dimensional harmonic oscillator. In the section 4,
the case of magnetic field is taken into account. Finally we trace
our final comments in the section 5.

\section{\label{sec1} Class of isospectral two-dimensional
position-dependent mass quantum systems}

In this section we present the approach which is capable to
generate a class of models with position-dependent masses from a
constant one, and which can also include the interaction with a
magnetic field. In this case the two-dimensional Schroedinger
equation is given by
\begin{equation}
-\frac{\hbar ^{2}}{2\,m_{0}}\,\nabla ^{2}\,\psi +\frac{\hbar }{i}\left(
\overrightarrow{\nabla }\cdot \overrightarrow{A}+2~\overrightarrow{A}\cdot
\overrightarrow{\nabla }\right) ~\psi +\left( V\left( x,y\right) +\frac{e^{2}%
}{2~m_{0}}\overrightarrow{A}^{2}\right) \,~\psi =E\,\psi .  \label{eq4a}
\end{equation}
where the constant $m_{0}$ is the mass of the particle in the original
system, $x$ and $y$ are the corresponding spatial coordinates, $p_{x}$ and $%
p_{y}$ are the respective momenta. Furthermore, we will work with
magnetic fields in the Coulomb gauge, where
$\overrightarrow{\nabla }\cdot \overrightarrow{A}=0$. Now,
performing a general variable transformation in the spatial
coordinates as
\begin{equation}
x=f(u,v),~y=g(u,v),  \label{eq4c}
\end{equation}
one gets
\begin{equation}
\,\partial _{u}\equiv \frac{\,\partial }{\,\partial u}=f_{u}\,\partial
_{x}+g_{u}\,\partial _{y},~\,\partial _{v}\equiv \frac{\,\partial }{%
\,\partial v}=f_{v}\,\partial _{x}+g_{v}\,\partial _{y},  \label{eq4f}
\end{equation}
which can be written as
\begin{equation}
\left(
\begin{array}{c}
\,\partial _{x} \\
\,\partial _{y}%
\end{array}
\right) =R^{-1}\,\left(
\begin{array}{c}
\,\partial _{u} \\
\,\partial _{v}%
\end{array}
\right) ,  \label{eq4h}
\end{equation}
where $R=\left(
\begin{array}{cc}
f_{u} & g_{u} \\
f_{v} & g_{v}%
\end{array}
\right) $ is the Jacobian matrix, and its Jacobian is $%
J=f_{u}~g_{v}-g_{u}~f_{v}$, from which it can be written that
\begin{equation}
\,\partial _{x}=\frac{\,\partial }{\,\partial x}=\frac{g_{v}}{J}\frac{%
\,\partial }{\,\partial u}-\frac{g_{u}}{J}\frac{\,\partial }{\,\partial v}%
,~\partial _{y}=\frac{\,\partial }{\,\partial y}=-\frac{f_{v}}{J}\frac{%
\,\partial }{\,\partial u}+\frac{f_{u}}{J}\frac{\,\partial }{\,\partial v}.
\label{eq4j}
\end{equation}
After straightforward calculations, one can obtain the following
expression for the Laplace operator in the transformed
coordinates,
\begin{eqnarray}
\nabla _{xy}^{2} &\equiv &\,\partial _{x}^{2}+\,\partial _{y}^{2}=\frac{1}{%
J^{2}}\left[ \left( g_{v}^{2}+f_{v}^{2}\right) \,\partial _{u}^{2}+\left(
g_{u}^{2}+f_{u}^{2}\right) \,\partial _{v}^{2}-2\left(
g_{u}g_{v}+f_{u}f_{v}\right) \,\partial _{uv}^{2}\right] +  \nonumber \\
&&+\frac{1}{J}\left[ g_{v}\,\partial _{u}\left( \frac{g_{v}}{J}\right)
-g_{u}\,\partial _{v}\left( \frac{g_{v}}{J}\right) +f_{v}\,\partial
_{u}\left( \frac{f_{v}}{J}\right) -f_{u}\,\partial _{v}\left( \frac{f_{v}}{J}%
\right) \right] \,\partial _{u}+  \nonumber \\
&&+\frac{1}{J}\left[ g_{u}\,\partial _{v}\left( \frac{g_{u}}{J}\right)
-g_{v}\,\partial _{u}\left( \frac{g_{u}}{J}\right) +f_{u}\,\partial
_{v}\left( \frac{f_{u}}{J}\right) -f_{v}\,\partial _{u}\left( \frac{f_{u}}{J}%
\right) \right] \,\partial _{v}.  \label{eq4o}
\end{eqnarray}
In order to achieve a system where we have a new Schroedinger-type equation
with a position dependent mass one must require that the crossed derivative
term $\,\partial _{uv}^{2}$, must vanish, leading to the restriction
\begin{equation}
g_{u}\,g_{v}+f_{u}~f_{v}=0.  \label{eq4p}
\end{equation}
Afterwards, we should also impose that
\begin{equation}
g_{v}^{2}+f_{v}^{2}=g_{u}^{2}+f_{u}^{2},  \label{eq4q}
\end{equation}
to guarantee that the mass term is the same in the both terms
$\,\partial _{u}^{2}$ and $\,\partial _{v}^{2}$. Note that the
equation (\ref{eq4p}) once solved leads to
\begin{equation}
f_{u}=-\frac{g_{u}\,g_{v}}{f_{v}},  \label{eq4r}
\end{equation}
provided that $f_{v}\neq 0$. By substituting (\ref{eq4r}) in (\ref{eq4q}),
one gets
\begin{equation}
g_{v}^{2}+f_{v}^{2}=g_{u}^{2}+g_{u}^{2}\left( \frac{g_{v}}{f_{v}}\right)
^{2}=g_{u}^{2}~\left( 1+\frac{g_{v}^{2}}{f_{v}^{2}}\right) =g_{u}^{2}~\left(
\frac{f_{v}^{2}+g_{v}^{2}}{f_{v}^{2}}\right) .  \label{eq4s}
\end{equation}
Thus, from (\ref{eq4s}) we conclude that
\begin{equation}
f_{v}^{2}=g_{u}^{2},~f_{v}=\pm g_{u}.  \label{eq4t}
\end{equation}
Substituting (\ref{eq4t}) in (\ref{eq4r}), one obtains
\begin{equation}
f_{u}=-\frac{g_{u}\,g_{v}}{\pm g_{u}}=\mp g_{v}.  \label{eq4u}
\end{equation}

In fact, in all the cases considered here, the terms linear in the
derivatives in the transformed Laplace operator (\ref{eq4o})
vanish. Moreover, the transformation will keep
$\overrightarrow{\nabla }\cdot \overrightarrow{A}=0$ \ for all the
cases considered. Thus the Schroedinger equation in those
transformed variables is written as
\begin{equation}
\left[ \frac{\hslash ^{2}}{2~M\left( u,v\right) }\left( \,\partial
_{u}^{2}+\,\partial _{v}^{2}\right) +V_{eff}\left( u,v\right) -\frac{i~\hbar
}{M\left( u,v\right) }\left( \overrightarrow{\tilde{A}}\cdot \overrightarrow{%
\nabla }\right) ~\right] \psi \left( u,v\right) =E~\psi \left( u,v\right) ,
\end{equation}

\noindent where
\begin{equation}
M\left( u,v\right) \equiv m_{0}\frac{J^{2}}{g_{u}^{2}+f_{u}^{2}}%
,~V_{eff}\left( u,v\right) \equiv V\left( f,g\right) +\frac{e^{2}}{2~M\left(
u,v\right) }\overrightarrow{\tilde{A}}^{2}\left( f,g\right) ,
\end{equation}

\noindent with $\overrightarrow{\tilde{A}}\equiv M\left( u,v\right) ~%
\overrightarrow{A}\left( f(u,v),g(u,v)\right) $.

As an illustrative example, we will first deal with
two-dimensional position-dependent systems which are isospectral
to the anisotropic harmonic oscillator and, then, with the ones
which are isospectral to the case of a isotropic harmonic
oscillator under the influence of homogeneous magnetic fields.

\section{2D PDM systems isospectral to the anisotropic oscillator}

As advertised, the first example which we will present here is the
one of the anisotropic oscillator with constant mass governed by
the equation
\begin{equation}
-\frac{\hbar ^{2}}{2~\,m_{0}}\,\nabla ^{2}\,\psi +\frac{1}{2}m_{0}\left(
\omega _{1}^{2}~x^{2}+\omega _{2}^{2}~y^{2}\right) \psi =E~\psi .
\end{equation}

\noindent In this case the eigenfunctions can be straightforwardly obtained,
and are given by

\begin{eqnarray}
\psi _{nm}(u,v) &=&\frac{1}{\sqrt{2^{(m+n+1)~}\pi ~n!~m!~X~Y}}~H_{m}\left(
\frac{\sqrt{2}\,f\left( u,v\right) }{X}\right) H_{n}\left( \frac{\sqrt{2}%
\,g\left( u,v\right) }{Y}\right) \times  \nonumber \\
&&\times exp\left[ -\left( \frac{f\left( u,v\right) }{X}\right) ^{2}-\left(
\frac{g\left( u,v\right) }{Y}\right) ^{2}\right] ,  \label{eq4tt3}
\end{eqnarray}

\noindent where $X=\sqrt{2\hbar /(m_{0}~\omega _{1})}$,
$Y=\sqrt{2\hbar /(m_{0}~\omega _{2})}$, $\ \omega _{1}\equiv q$
and $\omega _{2}\equiv p$.

At this point we could make a study of the eigenfunctions and
eigenvalues for the PDM systems which are isospectral to this one.
However, we prefer to construct the so-called $SU(2)$ coherent
states, as defined in \cite{1Chen2003}-\cite{Chen2004}. Those
states present the interesting feature of having their highest
probability density over a trajectory which corresponds to the
classical one when $\hbar\rightarrow 0$. Furthermore, as the
eigenstates, they are stationary wave-functions in contrast with
the usual coherent states. As we are going to see, they will lead
us to the conclusion that the isospectral PDM states will present
a behavior which is very similar to the one of their ``parent"
anisotropic harmonic oscillator. Those coherent states can be
written by using the definition introduced by Chen et al.
\cite{1Chen2003}-\cite{Chen2004}

\begin{equation}
\Phi (u,v,\tau )=\frac{1}{(1+|\tau |^{2})^{\frac{N}{2}}}\sum_{K=0}^{L}\left(
\begin{array}{c}
L \\
K%
\end{array}
\right) ^{1/2}\tau ^{K}\psi _{nm}(f(u,v),g(u,v)),  \label{eq4tt2}
\end{equation}

\noindent where the quantum numbers are defined as $n=p~K,$
$m=q(L-K)$, with $K=0,1,2,...,L$, $p$ and $q$ being integer
numbers. The complex parameter is such that $\tau=Ae^{i\,\phi }$,
where $\phi =\frac{\pi }{2}$, written in terms of polar
coordinates, is used in order to make the connection with the
classical trajectory.

From now on, we will devote this section to develop explicit
examples. As our first example of transformation functions, we
deal with the second degree polynomials like

\begin{equation}
f(u,v)=\frac{1}{2}a_{1}u^{2}+\frac{1}{2}b_{1}v^{2}+c_{1}u\,v+d_{1},~g(u,v)=%
\frac{1}{2}a_{2}u^{2}+\frac{1}{2}b_{2}v^{2}+c_{2}u\,v+d_{2}.  \label{eq4v}
\end{equation}

\noindent After imposing the restrictions (\ref{eq4p}) and (8), we
are led to

\begin{equation}
f(u,v)=\mp \frac{1}{2}c_{2}u^{2}\pm \frac{1}{2}c_{2}v^{2}+c_{1}u%
\,v+d_{1},~g(u,v)=\pm \frac{1}{2}c_{1}u^{2}\mp \frac{1}{2}%
c_{1}v^{2}+c_{2}u\,v+d_{2},  \label{eq4b1}
\end{equation}

\noindent where one can see that only four arbitrary constants are left, and
one have that

\begin{equation}
f_{u}^{2}+g_{u}^{2}=f_{v}^{2}+g_{v}^{2}=(c_{1}^{2}+c_{2}^{2})(u^{2}+v^{2}),
\label{eq4h1}
\end{equation}

\noindent and the Laplace operator becomes

\begin{equation}
\nabla _{uv}^{2}=\,\frac{1}{J^{2}}\left[ \left( g_{v}^{2}+f_{v}^{2}\right)
\,\partial _{u}^{2}+\left( g_{u}^{2}+f_{u}^{2}\right) \,\partial _{v}^{2}%
\right] =\frac{1}{(c_{1}^{2}+c_{2}^{2})(u^{^{2}}+v^{2})}(\partial
_{u}^{2}+\partial _{v}^{2}),  \label{eq4m1}
\end{equation}

\noindent from which one conclude that the position-dependent mass is given
by

\begin{equation}
M\left( u,v\right) = m_{0}(c_{1}^{2}+c_{2}^{2})(u^{^{2}}+v^{2}).
\end{equation}

\noindent In this case, the effective anisotropic potential looks like%
\begin{equation}
V\left( u,v\right) =\frac{m_{0}}{2}\left[ \left( \frac{c_{2}}{2}\left(
u^{2}-v^{2}\right) +c_{1}u~v+d_{1}\right) ^{2}\omega _{1}^{2}+\left( \frac{%
c_{1}}{2}\left( v^{2}-u^{2}\right) +c_{2}u~v+d_{2}\right) ^{2}\omega _{2}^{2}%
\right] ,
\end{equation}%
and the wavefunctions are written as%
\begin{eqnarray}
\psi _{nm}\left( u,v\right) &=&\frac{1}{\sqrt{2^{\left( m+n-1\right)
}m!n!P_{1}P_{2}}}H_{m}\left[ \frac{\sqrt{2}\left( \frac{c_{2}}{2}\left(
u^{2}-v^{2}\right) +c_{1}u~v+d_{1}\right) }{P_{1}}\right] \times  \nonumber
\\
&&H_{n}\left[ \frac{\sqrt{2}\left( \frac{c_{1}}{2}\left( v^{2}-u^{2}\right)
+c_{2}u~v+d_{2}\right) }{P_{2}}\right] \times \\
&&\exp \left[ -\left( \frac{\left( \frac{c_{2}}{2}\left( u^{2}-v^{2}\right)
+c_{1}u~v+d_{1}\right) }{P_{1}}\right) ^{2}\right] \times \\
&&\exp \left[ -\left( \frac{\left( \frac{c_{1}}{2}\left( v^{2}-u^{2}\right)
+c_{2}u~v+d_{2}\right) }{P_{2}}\right) ^{2}\right] .  \nonumber
\end{eqnarray}%
In the Figures 1 and 2 (we use $A=\hbar=m_0=1$ and $L=20$, when
generating the plots presented throughout this work), one can see
the behavior of the system in some typical situations. At this
point it is interesting to note that the transformation functions
can be chosen as the ones related to the parabolic cylindrical
coordinates in a plane, where $f=u~v$ and $g=\left(
u^{2}-v^{2}\right) /2$, which generate the following Laplace
operator
\begin{equation}
\nabla _{uv}^{2}=-~\frac{1}{(u^{^{2}}+v^{2})}(\partial
_{u}^{2}+\partial _{v}^{2}).
\end{equation}

\noindent This make us remember that one could work with a general
class of solutions as the ones coming from the orthogonal
coordinate systems, where the metric is diagonal. As examples we
will consider the case of the elliptic cylindrical coordinates on
a plane where

\begin{equation}
f(u,v)=a~\sinh (u)~\sin (v),~g(u,v)=a~\cosh (u)~\cos (v).  \label{eq4p1}
\end{equation}

\noindent In this case, the Laplace operator is written as
\begin{equation}
\nabla _{uv}^{2}=\,-\frac{2}{a^{2}[\cos (2v)-\cosh (2u)]}~(\partial
_{u}^{2}+\partial _{v}^{2}),
\end{equation}

\noindent with the mass $M\left( u,v\right) =m_{0~}a^{2}[\cos (2v)-\cosh
(2u)]/2$. The corresponding effective potential is now given by%
\begin{equation}
V(u,v)=\frac{a^{2}m_{0}}{2}\left[ \sinh \left( u^{2}\right) \sin \left(
v\right) ^{2}\omega _{1}^{2}+\cosh \left( u^{2}\right) \cos \left( v\right)
^{2}\omega _{2}^{2}\right]
\end{equation}%
The wavefuntions are written in this case as%
\begin{eqnarray}
\psi _{nm}\left( u,v\right) &=&\frac{1}{\sqrt{2^{\left( m+n-1\right)
}m!n!P_{1}P_{2}}}H_{m}\left[ \frac{\sqrt{2}\left( a~\sinh (u)~\sin
(v)\right) }{P_{1}}\right] \times  \nonumber \\
&&H_{n}\left[ \frac{\sqrt{2}\left( a~\cosh (u)~\cos (v)\right) }{P_{2}}%
\right] \exp \left[ -\left( \frac{a~\sinh (u)~\sin (v)}{P_{1}}\right) ^{2}%
\right] \times \\
&&\exp \left[ -\left( \frac{a~\cosh (u)~\cos (v)}{P_{2}}\right) ^{2}\right] ,
\nonumber
\end{eqnarray}

\noindent whose typical behavior appears in the Figure 3.

Now, in the bispherical coordinates on a plane, the
transformations are

\begin{equation}
f(u,v)=\frac{a~\sinh (u)}{\cosh (u)-\cos (v)},~g(u,v)=\frac{a~\sin (v)}{%
\cosh (u)-\cos (v)},
\end{equation}%
and the corresponding Laplace operator looks like
\begin{equation}
\nabla _{uv}^{2}=\,\frac{[\cos (v)-\cosh (u)]^{2}}{a^{2}}~(\partial
_{u}^{2}+\partial _{v}^{2}),
\end{equation}

\noindent and the position-dependent mass will be given by $M\left(
u,v\right) =m_{0}~a^{2}/[\cos (v)-\cosh (u)]^{2}$. In this last example of
anisotropic system, the effective potential happens to be%
\begin{equation}
V(u,v)=\frac{a^{2}m_{0}}{2}\left[ \left( \frac{a~\sinh (u)}{\cosh (u)-\cos
(v)}\right) ^{2}\omega _{1}^{2}+\left( \frac{a~\sin (v)}{\cosh (u)-\cos (v)}%
\right) ^{2}\omega _{2}^{2}\right] ,
\end{equation}

\noindent and the wavefunctions are correspondingly given by%
\begin{eqnarray}
\psi _{nm}\left( u,v\right) &=&\frac{1}{\sqrt{2^{\left( m+n-1\right)
}m!n!P_{1}P_{2}}}H_{m}\left[ \frac{\sqrt{2}}{P_{1}}\left( \frac{a~\sinh (u)}{%
\cosh (u)-\cos (v)}\right) \right] \times  \nonumber \\
&&H_{n}\left[ \frac{\sqrt{2}}{P_{2}}\left( \frac{a~\sin (v)}{\cosh (u)-\cos
(v)}\right) \right] \exp \left\{ -\left[ \frac{1}{P_{1}}\left( \frac{a~\sinh
(u)}{\cosh (u)-\cos (v)}\right) \right] ^{2}\right\} \times \\
&&\exp \left\{ -\left[ \frac{1}{P_{2}}\left( \frac{a~\sin (v)}{\cosh
(u)-\cos (v)}\right) \right] ^{2}\right\} ,  \nonumber
\end{eqnarray}

\noindent whose typical behavior appears in the Figure 4.

\section{\label{sec2} The case of interaction with a magnetic field}

In this section, we first construct the $SU(2)$ coherent state for
the case of a particle with constant mass in the presence of a
homogeneous magnetic field which, as far as we know, was not yet
obtained in the literature. Then, we proceed by using that result
in order to achieve the $SU(2)$ coherent states for the PDM
particles under the presence of some non-homogeneous magnetic
fields. All the results will be obtained in the Coulomb gauge.
Starting with the Hamiltonian

\begin{equation}
H=\frac{1}{2~m_{0}}\overrightarrow{p}^{2}-\frac{e}{2~m_{0}}\left(
\overrightarrow{A}.\overrightarrow{p}+\overrightarrow{p}.\overrightarrow{A}%
\right) +\frac{e^{2}}{2~m_{0}}\overrightarrow{A}^{2}+\frac{1}{2}m_{0}\omega
^{2}~\left( x^{2}+~y^{2}\right) ,  \label{eq4rr1}
\end{equation}%
where $e$ is the electrical charge and $\overrightarrow{A}(x,y)$ is the
vector potential.

Here we choose to work with a uniform magnetic field along the
direction $z$, which is written as
$\overrightarrow{B}=B_{0\;}\hat{z}$. In the so-called symmetrical
gauge it is written as
\begin{equation}
\overrightarrow{A}=\frac{B_{0}}{2}\left( -y~\hat{\imath}+x~\hat{\jmath}%
\right) ,  \label{eq4tt}
\end{equation}%
Thus, beginning with the classical Hamiltonian defined in
(\ref{eq4rr1}) and quantizing it, one arrives at the following
Schroedinger equation

\begin{equation}
-\frac{\hbar ^{2}}{2~\,m_{0}}\,\nabla ^{2}\,\psi +\frac{i~\hbar ~B_{0}}{%
2~m_{0}}\left( x~\partial _{y}-y~~\partial _{x}\right) \psi +\frac{1}{2}%
m_{0}\Omega ^{2}\left( x^{2}+~y^{2}\right) \psi =E\psi .  \label{eq4uu1}
\end{equation}

\noindent with $\Omega\equiv \omega_0^2+\frac{e^{2}B_{0}^{2}}{8~m_{0}}$.

In order to deal with the case of a constant mass in the presence
of a homogeneous magnetic field, one can follow two alternative
routes. One way is to write the system in polar coordinates and
the another is keeping the cartesian coordinates and performing
some convenient time-dependent transformations. This last is the
one we will follow here \cite{pra89}. For this, we can start with
the time-dependent Schroedinger equation

\begin{equation}
-\frac{\hbar ^{2}}{2\,m_{0}}\nabla ^{2}\sigma +\frac{i\,\hbar \,B_{0}\,e}{%
2\,m_{0}}\left( x\,\partial y-y\partial x\right) \sigma +\frac{1}{\,m_{0}}%
U_{eff}\,\sigma =i\,\hbar \,\frac{\partial \sigma }{\partial t},
\label{eq3vv2}
\end{equation}

\noindent such that $\sigma =e^{-(i/\hbar )Et}\chi (x,y)$. Then, we perform
the time-dependent rotation

\begin{equation}
\left(
\begin{array}{c}
X_{1} \\
Y_{1}%
\end{array}%
\right) =\left(
\begin{array}{cc}
cos\alpha (t) & sen\alpha (t) \\
-\,sen\alpha (t) & cos\alpha (t)%
\end{array}%
\right) \left(
\begin{array}{c}
x \\
y%
\end{array}%
\right) ,  \label{eq4aa1}
\end{equation}

\noindent which, after some manipulations \cite{pra89}, and
choosing $\alpha$ to be given by

\begin{equation}
\alpha (T)=-\frac{e\,B_{0}}{2\,m_{0}}~T+c,  \label{eq4uu2}
\end{equation}

\noindent where $c$ is an arbitrary integration constant. One can finish
with an effective isotropic two-dimensional harmonic oscillator such that

\begin{equation}
-\frac{\hbar ^{2}}{2~m_{0}}\nabla ^{2}\sigma +\left[ \left( \frac{%
e^{2}B_{0}^{2}}{8~m_{0}}+\frac{1}{2}m_{0}~\omega ^{2}\right) \left(
X_{1}^{2}+Y_{1}^{2}\right) \right] \,\sigma =i\,\hbar \,\frac{\partial
\sigma }{\partial T}.  \label{eq4xx1}
\end{equation}

This allow us to map the original problem into one where the differential
equation is given by

\begin{equation}
-\frac{\hbar ^{2}}{2M_{0}}\,\nabla ^{2}\chi +U_{eff}\,\chi =E\,\chi ,
\label{eq3y1}
\end{equation}

\noindent where $\chi =\chi (X_{1},Y_{1})$. Finally, using the usual
variables separation procedure, one can arrive at the expression

\begin{eqnarray}
\chi _{nm}(X_{1},Y_{1}) &=&\phi (X_{1})~\psi (Y_{1})=  \nonumber \\
&=&\phi \left[ cos(\alpha )\,x+\,sen(\alpha )\,y\right] \,\psi \left[
-sen(\alpha )\,x\,+\,cos(\alpha )\,y\right] =  \nonumber \\
&=&\frac{(M\left[ cos(\alpha )\,x+\,sen(\alpha )~\,y,-sen(\alpha
)\,x\,+\,cos(\alpha )\,~y\right] )^{-1}}{R~\sqrt{2^{m+n-1}\pi n!m!}}~\times
\nonumber \\
&\times &H_{m}\left( \frac{\sqrt{2}[cos(\alpha )\,x+\,sen(\alpha )~\,y]}{R}%
\right) \times  \label{wavemag} \\
&&~exp\left[ -\left( \frac{cos(\alpha )\,x+\,sen(\alpha )~\,y}{R}\right) ^{2}%
\right] \times  \nonumber \\
&\times &~H_{n}\left( \frac{\sqrt{2}[-sen(\alpha )\,x\,+\,cos(\alpha )\,~y]}{%
R}\right) \times  \nonumber \\
&\times &exp\left[ -\left( \frac{-sen(\alpha )\,x\,+\,cos(\alpha )\,~y}{R}%
\right) ^{2}\right]  \nonumber
\end{eqnarray}

Finally, by using the above wave-functions, the $SU(2)$ coherent
states can be easily written. Their general aspect happens to be
the same of the isotropic harmonic oscillator as presented in the
left plot of Figure 2.

Now, we can study the cases of PDM particles in the presence of
magnetic fields. Following the same lines developed in the
previous section, we make the variables transformation $x=f(u,v)$
and $y=g(u,v)$ in the Schroedinger equation and, since we already
now the corresponding transformation in the Laplace operator, we
only need to verify the impact of the transformation
over the linear differential operator coming from $\overrightarrow{A}.%
\overrightarrow{p}$, which comes to be

\begin{equation}
x\,\partial _{y}-y\,\partial _{x}=\frac{1}{J}\left\{
[~f(u,v)~\,f_{u}+g(u,v)~\,g_{u}~]~\partial
_{v}-[~f(u,v)\,f_{v}+g(u,v)\,g_{v}~]~\partial _{u}\right\} .
\end{equation}

Then, we finish with the transformed Schroedinger equation in the
spatial variables $u$ and $v$ appearing as
\begin{eqnarray}
&&-\frac{\hbar ^{2}}{2\,M(u,v)}\,\nabla _{uv}^{2}\psi +\frac{i\hbar B_{0}}{%
2~M(u,v)}\left[ R(u,v)\frac{\partial \psi }{\partial u}+S(u,v)\frac{\partial
\psi }{\partial v}\right] +  \nonumber \\
&&+\left\{\frac{1}{2}m_{0}\Omega ^{2}\left[ ~f(u,v)^{2}+g(u,v)^{2}\right]
\right\} \psi =E\psi .
\end{eqnarray}

The magnetic field will be then given by

\begin{equation}
\overrightarrow{B}=\overrightarrow{\nabla }\times \overrightarrow{A}=\left(
\begin{array}{ccc}
\overrightarrow{i} & \overrightarrow{j} & \overrightarrow{k} \\
\frac{\partial }{\partial _{u}} & \frac{\partial }{\partial _{v}} & \frac{%
\partial }{\partial _{w}} \\
R & S & 0%
\end{array}
\right) =\left( \frac{\partial S}{\partial u}-\frac{\partial R}{\partial v}%
\right) \overrightarrow{k}.
\end{equation}

Again, the first example to be analyzed is the one defined in Equation (\ref%
{eq4b1}). In this case we have that

\begin{eqnarray}
R(u,v) &=&-\frac{1}{2}\left( c_{1}^{2}+c_{2}^{2}\right) \left(
v^{3}+u^{2}\,v\right) -\left( c_{1}d_{1}+c_{2}d_{2}\right)
u-\left(
c_{1}d_{2}-c_{2}d_{1}\right) v,  \nonumber \\
S(u,v) &=&\frac{1}{2}\left( c_{1}^{2}+c_{2}^{2}\right) \left(
u^{3}+u\,v^{2}\right) +\left( c_{2}d_{1}-c_{1}d_{2}\right)
u+\left( c_{1}d_{1}+c_{2}d_{2}\right) v.  \label{eq4w2}
\end{eqnarray}

\noindent Here, we will have a particle with mass $M\left( u,v\right) =
m_{0}(c_{1}^{2}+c_{2}^{2})(u^{^{2}}+v^{2})$ moving in the presence of an
axially symmetric field with a quadratically growing intensity,

\begin{equation}
\overrightarrow{B}=B_{0}~(c_{1}^{2}+c_{2}^{2})(u^{2}+v^{2})~\overrightarrow{k%
}.  \label{eq4w3}
\end{equation}

Choosing the parameters $m_0=d_{1}=d_{2}=c_{1}=0$ and $c_{2}=1$, we restrict
ourselves to the case of parabolic cylinder coordinates where

\begin{equation}
M(u,v)=u^{2}+v^{2}\,,~R(u,v)=-\frac{1}{2}v~(u^{2}+v^{2}),~S(u,v)=\frac{1}{2}%
u~(u^{2}+v^{2}),
\end{equation}

\noindent and the potential governing the system is%
\begin{eqnarray}
V\left( u,v\right) &=&\frac{m_{0}~\omega ^{2}}{2}\left[ \frac{1}{4}\left(
u^{2}-v^{2}\right) ^{2}+\left( u~v\right) ^{2}\right] +  \nonumber \\
&& \\
&&-e^{2}B_{0}\sqrt{u^{2}+v^{2}},  \nonumber
\end{eqnarray}

\noindent and the wavefunctions are%
\begin{eqnarray}
\psi _{nm}(u,v) &=&\frac{1}{R~\sqrt{2^{m+n-1}\pi n!m!}}~H_{m}\left( \frac{%
\sqrt{2}[\frac{\cos (\alpha )}{2}\left( u^{2}-v^{2}\right) +\,\sin (\alpha
)~\,u~v]}{R}\right)  \nonumber \\
&&\times H_{n}\left( \frac{\sqrt{2}[-\frac{\sin (\alpha )}{2}\left(
u^{2}-v^{2}\right) +\cos (\alpha )~\,u~v]}{R}\right) \\
&&\times exp\left[ -\left( \frac{u^{2}-v^{2}}{2~R}\right) ^{2}-\left( \frac{%
u~v}{R}\right) ^{2}\right] ,  \nonumber
\end{eqnarray}

\noindent which, after plotting the probability density of the
corresponding SU(2) coherent state, presents the profile which is
very similar to the one appearing in the Figure 1.

\noindent \qquad Now, in the case of elliptical cylindrical
coordinates one have
\begin{equation}
R(v)=a^{2}~\cos (v)~\sin (v),~S(u)=a^{2}~\cosh (u)~\sinh (u).  \label{eq4ff3}
\end{equation}

\noindent and the magnetic field is given by

\begin{equation}
\overrightarrow{B}=~a^{2}[\cosh (2u)-\cos (2v)]~\overrightarrow{k}.
\end{equation}

The potential under which the charged particle is moving is%
\begin{eqnarray}
V\left( u,v\right) &=&\frac{m_{0}~a^{2}~\omega ^{2}}{2}\left[ \sinh \left(
u^{2}\right) \sin \left( v\right) ^{2}+\cosh \left( u^{2}\right) \cos \left(
v\right) ^{2}\right] +  \nonumber \\
&& \\
&&-e^{2}B_{0}~\frac{\sqrt{\left[ \cosh (4~u)-\cos (4~v)\right] }}{2~\left[
\cos (2v)-\cosh (2~u)\right] },  \nonumber
\end{eqnarray}

\noindent and the wavefunctions can be straightforwardly obtained
through direct substitution of the transformation functions in the
expression of the wavefunctions (\ref{wavemag}). After that, one
can plot the probability density, which present the same general
behavior as appearing in the Figure 3. There, one can see a
sequence of regions where the highest probability density is over
a closed curve, which reflects the periodicity of the
transformation functions.

In the third and last example, we will deal with the bipolar
coordinates. In this case the Laplace operator is written as
$\nabla _{uv}^{2}=\,\frac{[\cos (v)-\cosh
(u)]^{2}}{a^{2}}~(\partial _{u}^{2}+\partial _{v}^{2})$, and we
get
\begin{equation}
R(u,v)=\frac{a^{2}~\cosh (u)~\sin (v)}{[\cos (v)-\cosh (u)]^{2}},~S(u,v)=-%
\frac{a^{2}~\cos (v)~\sinh (u)}{[\cos (v)-\cosh (u)]^{2}}.  \label{eq4l3}
\end{equation}

\noindent the corresponding magnetic field looks like

\begin{equation}
\overrightarrow{B}=\frac{2~a^{2}}{[\cos (v)-\cosh (u)]^{2}}~\overrightarrow{k%
},
\end{equation}%
\newline

\noindent and, in this last example, the potential is written as%
\begin{equation}
V(u,v)=\frac{a^{2}m_{0}~\omega ^{2}}{2}\frac{\left( \sinh (u)^{2}+\sin
(v)^{2}\right) }{\left( \cosh (u)-\cos (v)\right) ^{2}}-\frac{e^{2}B_{0}}{%
4~m_{0}}~\sqrt{\cosh (u)^{2}-\cos (v)^{2}}.
\end{equation}

Once more, after constructing the wave-function for the $SU(2)$ coherent
state, one can see that we arrive at a Figure which is very similar to the
case (a) of the Figure 4 below.\bigskip

\section{\label{sec3}Final remarks}

In this work we explored a method for generating exactly solvable
position-dependent mass particle in the present of some
potentials. This approach was recently introduced in the case of
one-dimensional systems \cite{juliano1}. Here we see that the
extension for higher dimensions is not trivial and allows one to
analyze complex and interesting new systems. Moreover, we conduce
our study by computing the so called $SU(2) $ coherent states,
which are specially interesting wave-packets which
present their maximum of probability over the classical trajectory \cite%
{juliano2,1Chen2003,3Chen2003,Chen2004}. As particular examples of
variable transformation, we used a polynomial example as well as
the cases of the cylindrical elliptical and the bipolar
coordinates.

In the case of a charged particle under the effect of pure
homogeneous external magnetic field, despite the fact that the we
are dealing with time-dependent wave-function solutions, it was
verified that the corresponding $SU(2)$ coherent states do not
present dependence in the time variable. In that case, the maximum
of the probability density is over a circle. Then, performing the
change in the spatial variables, we got those states for the case
of some particular spatially dependent masses and non-homogenous
magnetic fields. In the case where the transformation used was
that of polynomial form, it was observed that when the zero degree
coefficient of the polynomial  used in the transformation was
non-zero, the maximum of probability splits into two ones which
become more and more  distant from each other when that parameter
increases. Furthermore, in the case of the other two
transformations used  here, the probability density was naturally
periodic as a direct consequence of the periodicity of the
transformation.

\bigskip
\textbf{Acknowledgements: }The authors thanks to CNPq and CAPES
for partial financial support. J. A. O. thanks the DFQ of UNESP,
Campus de Guaratinguet\'a, where this work was carried out. This
work was partially done during a visit (ASD) within the Associate
Scheme of the Abdus Salam ICTP.

\newpage

\begin{figure}[tbp]
\centering
\includegraphics[width=5.5cm]{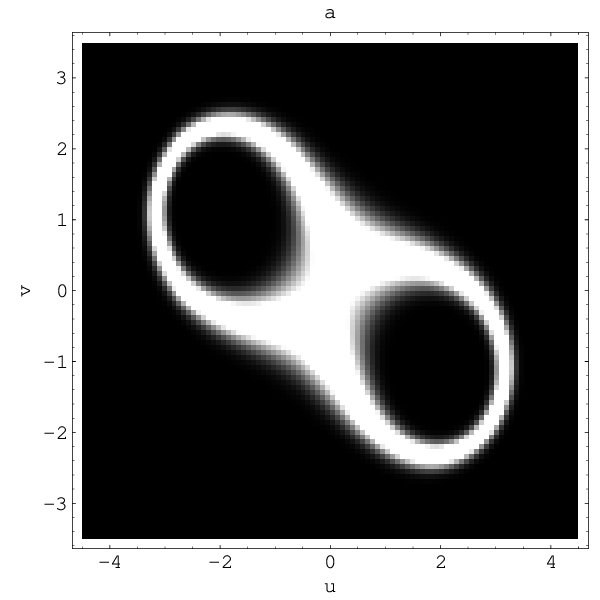} %
\includegraphics[width=5.5cm]{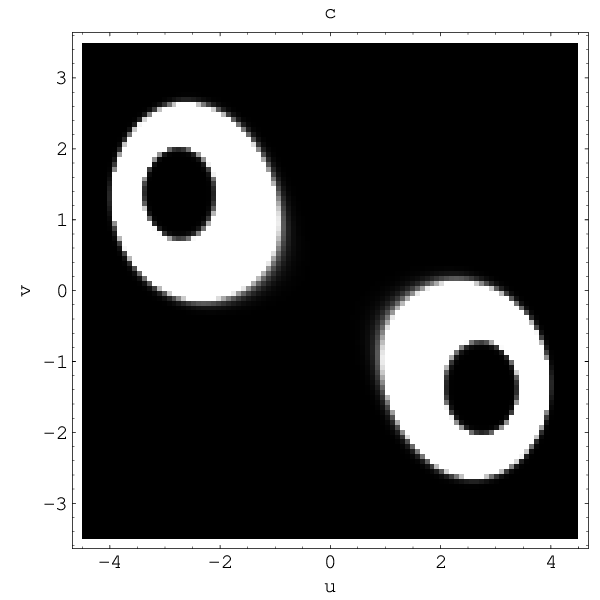}
\caption{\textit{Probability density for the case of polynomial
solutions. Here the parameters used are $a=c_1=c_2=1$ and: a) $d_{2}=4$ and c) $%
d_{2}=7$.}}
\end{figure}

\begin{figure}[tbp]
\centering
\includegraphics[width=5.5cm]{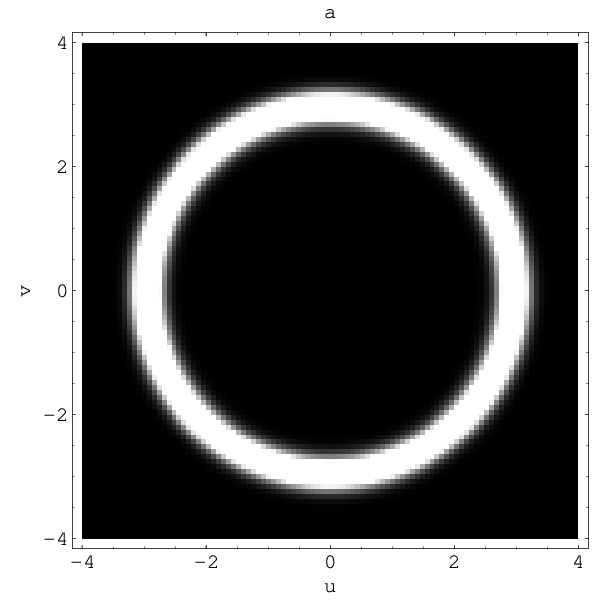} %
\includegraphics[width=5.5cm]{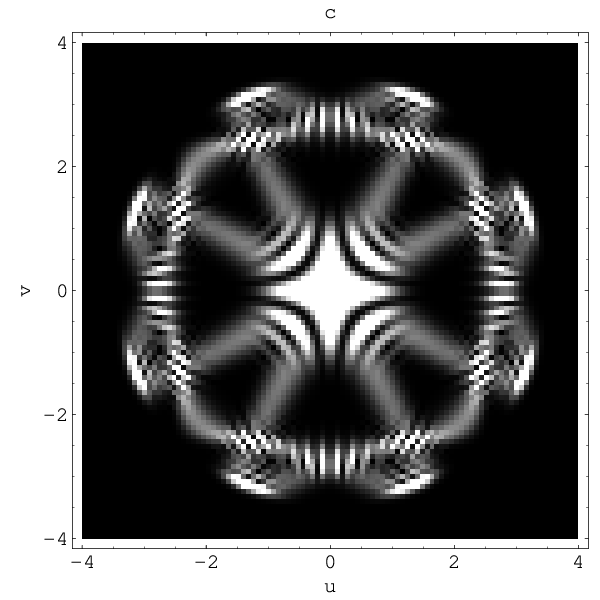}
\caption{\textit{Probability density for the case of polynomial
solutions. Here the parameters used are: $d_{1}=d_{2}=c_{1}=0$,
$c_{2}=1$
and the frequencies are the following: a) $\protect\omega _{1}=1$ and $\protect\omega %
_{2}=1$ and c) $\protect\omega _{1}=2$ and $\protect\omega
_{2}=3$.}} \label{para_cilin_uv}
\end{figure}

\begin{figure}[tbp]
\centering
\includegraphics[width=5.5cm]{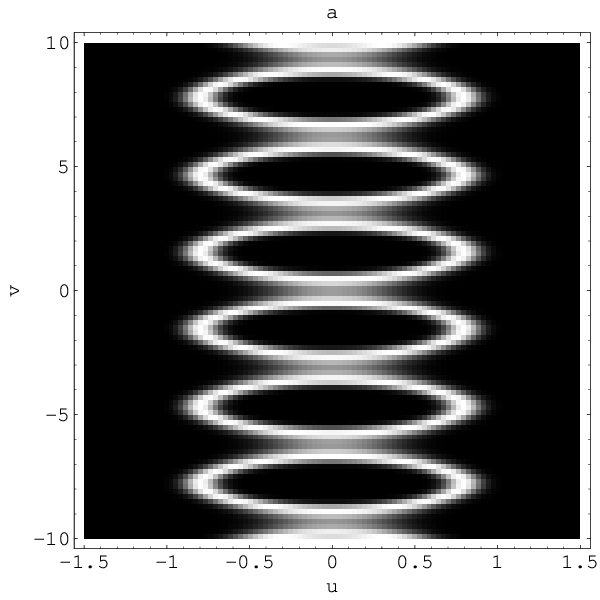}
\caption{\textit{Probability density for the case of elliptical
cylindrical coordinates. In this case the frequencies were: a)
$\protect\omega _{1}=1$ and $\protect\omega _{2}=1$.}}
\end{figure}

\begin{figure}[tbp]
\centering
\includegraphics[width=5.5cm]{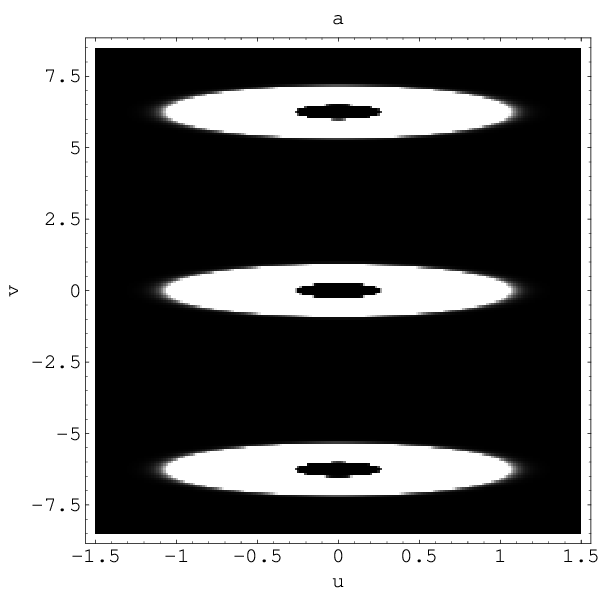} %
\includegraphics[width=5.5cm]{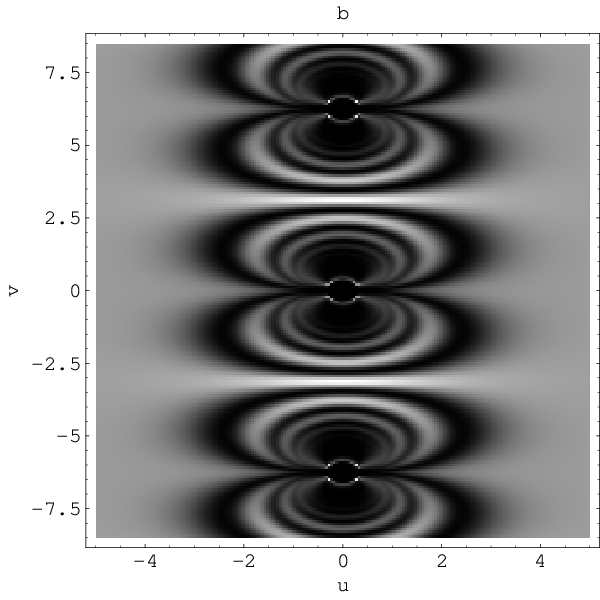} %
\includegraphics[width=5.5cm]{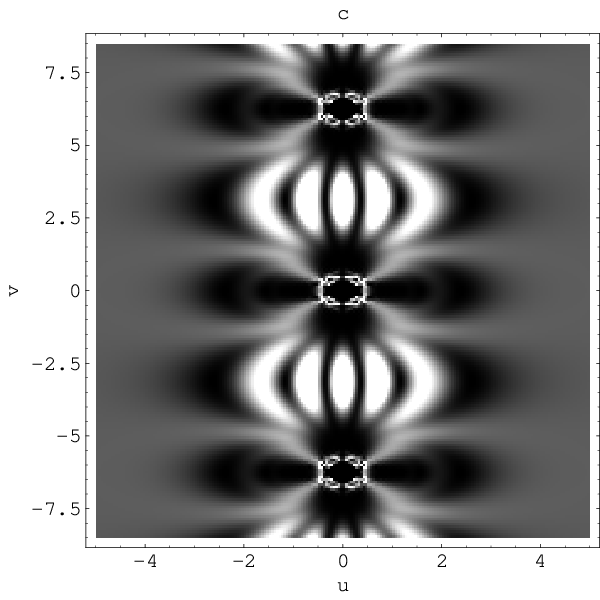}
\caption{\textit{Probability density for the case of bipolar coordinates. The used frequencies were: a) $%
\protect\omega _{1}=1$ and $\protect\omega _{2}=1$, b)
$\protect\omega _{1}=1$
and $\protect\omega _{2}=2$ and c) $\protect\omega _{1}=2$ and $\protect\omega %
_{2}=3$.}}
\label{cilin_bipo}
\end{figure}

\end{document}